\documentclass[prb,twocolumn,superscriptaddress,showpacs,amssymb]{revtex4}
\usepackage{graphicx}

\begin{document}
\title{Depth-dependent critical behavior in V${}_2$H}
\author{Charo I. \surname{Del Genio}}
	\affiliation{University of Houston, Department of Physics, 617 Science \& Research 1, 4800 Calhoun Rd, Houston, TX 77204-5005}
	\affiliation{Texas Center for Superconductivity, Houston, TX 77204}
\author{Johann Trenkler}
	\altaffiliation[Currently at ]{Carl Zeiss SMT AG, Lithography Optics Division, Rudolf-Eber-Stra\ss e~2, D-73447~Oberkochen, Germany}
	\affiliation{University of Houston, Department of Physics, 617 Science \& Research 1, 4800 Calhoun Rd, Houston, TX 77204-5005}
	\affiliation{Max Planck Institut f\"{u}r Metallforschung, D-70569 Stuttgart, Germany}
\author{Kevin E. \surname{Bassler}}
	\affiliation{University of Houston, Department of Physics, 617 Science \& Research 1, 4800 Calhoun Rd, Houston, TX 77204-5005}
	\affiliation{Texas Center for Superconductivity, Houston, TX 77204}
\author{Peter Wochner}
	\affiliation{Max Planck Institut f\"{u}r Metallforschung, D-70569 Stuttgart, Germany}
\author{Dean R. \surname{Haeffner}}
	\affiliation{Advanced Photon Source, Argonne National Laboratory, Argonne, IL 60439-4815}
\author{George F. \surname{Reiter}}
	\affiliation{University of Houston, Department of Physics, 617 Science \& Research 1, 4800 Calhoun Rd, Houston, TX 77204-5005}
\author{Jianming Bai}
	\affiliation{Oak Ridge National Laboratory, Oak Ridge, TN 37831}
\author{Simon C. \surname{Moss}}
	\affiliation{University of Houston, Department of Physics, 617 Science \& Research 1, 4800 Calhoun Rd, Houston, TX 77204-5005}
	\affiliation{Texas Center for Superconductivity, Houston, TX 77204}

\date{\today}

\begin{abstract}
Using X-ray diffuse scattering, we investigate the critical behavior of an
order-disorder phase transition in a defective ``skin-layer'' of V$_2$H.
In the skin-layer, there exist walls of dislocation lines  
oriented normal to the
surface. 
The density of dislocation lines within a wall decreases continuously with depth.
We find that, because of this inhomogeneous distribution of defects, 
the transition effectively occurs at a depth-dependent local critical 
temperature.
A depth-dependent scaling law is proposed to describe
the corresponding 
critical ordering behavior.
\end{abstract}


\maketitle

\section{Introduction}

Structural defects exist in almost all real crystalline solids.
Therefore, in order to understand structural phase transitions,
it is crucial to understand the influence that
defects can have on ordering behavior.
It has been shown that defects, through their accompanying
strain fields, can change the nature of phase transitions,
including their universal critical properties~\cite{Dub79,Kor99}.
Defects have also been shown to be responsible for the appearance
of the so-called ``central peak'' in diffuse X-ray or neutron scattering
and the related ``two-length scale'' phenomena observed
in an number of experimental systems~\cite{And86,Mcm90,Geh93,Thu93,Hir94,Hir95,Neu95,Cow96,Geh96,Rut97,Wan98,Kor99}.
These previous
studies have implicitly assumed that defects are homogeneously distributed
in the material, at least in the region of the crystal being studied,
even if that region is only a ``skin-layer'', i.e., a near-surface region in
which the defect density is known to be different than in the bulk.
In real systems, however, defects are often caused by surface
treatments. In this case, they can be inhomogeneously
distributed, occurring mostly in a skin-layer with a
density that continuously decays into the bulk over several microns~\cite{Tre98,Wan98,Tre01}.
Such an inhomogeneous distribution of defects complicates
the ordering behavior of many real crystals.
In this letter, using diffuse X-ray scattering in both reflection
and transmission geometries, we analyze the depth-dependent
critical behavior of the
structural ordering of a crystal with this type of
inhomogeneous defect distribution.

Divanadium hydride (V${}_2$H) is an interstitial alloy
that undergoes a structural phase transition between an ordered monoclinic
phase $\beta_1$ and a disordered body centered tetragonal phase $\beta_2$
as temperature is increased (for the phase diagram see Ref.~\onlinecite{Sch79}).
In the crystal we study, defects occur almost exclusively in a skin-layer that
extends several $\mu$m below the surface. In the skin-layer,
there exist walls of dislocation lines oriented normal to the surface~\cite{Tre01}.
The density of dislocation lines in the walls decreases
with depth. As we will see, the
character of the structural transition in the crystal
can change radically with the depth
at which it takes place. In the bulk material it
is first order\cite{Tre99}, but it becomes continuous in the skin-layer and has a
critical temperature and critical properties that depend on depth.
We propose a modification
of the scaling law for the inverse correlation length that accounts
for its depth-dependence and allows us to
treat the scattering measurements at different depths
in a unified framework.

\section{Experimental setup and results}

For our experiments we used a thin plate (0.96~mm) of a vanadium
single crystal loaded with
purified hydrogen, so that it had a
bulk concentration ratio
of $c_\mathrm{H}/c_\mathrm{V}=0.525\pm0.005$. 
For sample preparation and analysis 
see Refs.~\onlinecite{Tre98} and~\onlinecite{Tre00}.
We performed X-ray experiments on the crystal
in both reflection and
transmission geometry.
The reflection experiments were performed with MoK${}_{\alpha 1}$ X-rays
at a rotating anode source and at several energies 
on X14A at the NSLS at Brookhaven National Laboratory.
The high energy transmission experiments 
were carried out 
with 44.1~keV X-rays 
at the
undulator beamline SRI-CAT, 1-ID, at the APS at Argonne National 
Laboratory~\cite{Tre99}.
In all the experiments we confirmed the absence of higher harmonic contamination.
Since we earlier observed two length scales in this 
crystal~\cite{Tre98} we used
the different scattering geometries and energy ranges 
in order to detect separately
the influences arising from the bulk and the skin layer. 
In all cases the sample was
mounted in a strain-free manner in a vacuum of $\sim 10^{-4}$~torr. 
The temperature
fluctuations of the entire setup were less than 0.05~K at $T>443$~K.

We have four indications of a
defective near-surface layer on our sample:
(1) A hydrogen and oxygen gradient measured by high resolution elastic
recoil detection analysis (HERDA) in the first 150~\AA{}~\cite{Tre00};
(2) an oxygen gradient measured by secondary neutral mass spectroscopy
(SNMS) in the first 150--200~\AA{}~\cite{Tre00};
(3) the decay of the mosaic spread with depth~\cite{Tre98,Tre01};
(4) a larger $d$-spacing in the near-surface region than in the bulk~\cite{Tre98}.
HERDA shows that the hydrogen content increases with 
depth until the bulk concentration is reached at a
depth of about 150~\AA{}.
In our most surface sensitive 
experiment we have used 9~keV
X-rays and low momentum transfers to measure the influence 
of the upper 150~\AA{}
on the scattering; 
this contribution appears to be about 0.2~\%, as deduced from the
fraction $R$ of the intensity diffracted down to a depth $d$ in the sample. 
In a symmetric
scattering geometry, $R$ is given by
\begin{equation}
 R=1-\mathrm{e}^{-2\frac{\mu d}{\sin \Theta}}\:,
\end{equation}
where $\mu$ is the linear absorption coefficient of X-rays and $\Theta$
is the Bragg angle~\cite{LSch87}. The only defects penetrating up to several
microns in our sample are thus the walls of dislocations responsible for the mosaic spread.
We note that there is a depth-dependent stress field associated with this decay,
caused by walls of dislocation lines~\cite{Tre01}.
\begin{figure}
 \centering
\includegraphics[width=0.45\textwidth]{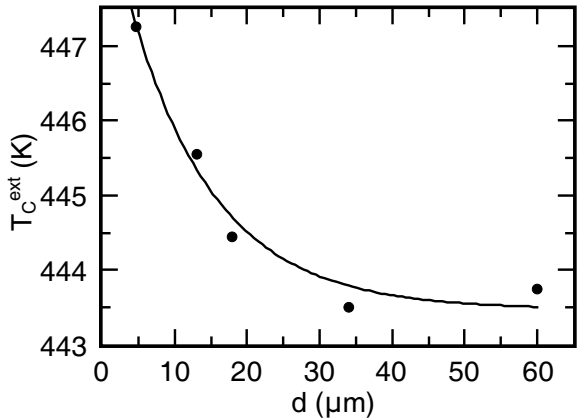}
\caption{\label{Fig1}Critical temperature $T_C^{ext}$ versus depth $d$.}
\end{figure}

We focus here on the depth dependence of the critical behavior in the
vicinity of the $\beta_1$-$\beta_2$ order-disorder transition in the
several micron thick skin-layer.
The correlation length
$\xi=1/\kappa$ for $T\geqslant T_C^{ext}$ at a constant
depth is the inverse of the half width at half maximum of the critical
diffuse scattering (CDS) profiles, and could be reliably determined only
through a fit involving a convolution of the measured resolution function
and the CDS profiles. The critical temperature $T_C^{ext}\left( d\right)$, whose value depends
on the depth $d$ (see Fig.~\ref{Fig1}), is defined here as the
extrapolated temperature at which the full width at half maximum
reaches 0.
Note that $T_C^{ext}\left( d\right)$ is an ``integrated'' quantitity
that depends on the scattering from the entire skin-layer down to depth $d$.
The fact that $T_C^{ext}\left( d\right)$ depends on depth, presumably implies that there is a 
real, local critical temperature $\hat{T}_C\left( d\right)$ that also depends on depth.
Note also that the concept of a local critical temperature is well-defined
only within a region smaller than the local correlation length. In other words,
for any particular depth $d^\ast$, when the temperature approaches $T_C\left( d^\ast\right)$
the ordering process will happen in a layer around that depth not thicker than the local correlation
length.

The high energy transmission experiments indicated a strong
first order phase transition in the bulk, evidenced by
a strong drop of the $( 0\quad\!5/2\quad\!\bar 5/2 )$ superstructure intensity by a
factor of more than 400 at $T_C^{ext}$ (see Ref.~\onlinecite{Tre99}). This is associated with a transition
width of~$\approx 0.3$~K, together with an abrupt broadening of the intensity
profile at the superstructure position.

However, in the skin layer the transition is continuous and the long-range 
order parameter exponent $\beta$ can be determined from the integrated Bragg
intensities of superstructure reflections $I$:
\begin{equation}
 I\propto\Phi^2=-B\left( \frac{T}{T_C^{ext}}-1\right) ^{2\beta}\:,
\end{equation}
where $\Phi$ is the Bragg-Williams order parameter, and $B$ a constant.
From the corrected intensities for the 
$( 0\quad\!5/2\quad\!\bar 5/2 )$ and $( 0\quad\!7/2\quad\!\bar 7/2 )$ 
superstructure reflections, after excluding a small two-phase region,
we obtained a value of $\beta=0.18\pm0.02$ (see Ref.~\onlinecite{Tre00}), 
treating $T_C^{ext}$ as a fit parameter as, e.g., in Ref.~\onlinecite{BSch87}.
\begin{figure}
 \centering
\includegraphics[width=0.45\textwidth]{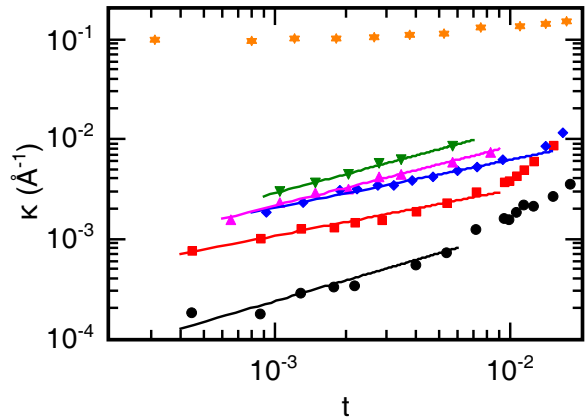}
\caption{\label{Fig2}(Color online) Inverse correlation length $\kappa$ versus reduced temperature
$t=\frac{T}{T_C^{ext}}-1$ for different depths. The black circles
correspond to 1.6~$\mu$m, the red squares to 13.1~$\mu$m, the blue diamonds to
18.4~$\mu$m, the pink upward triangles to 25~$\mu$m, the green downward triangles
to 34~$\mu$m and the orange stars to the bulk.}
\end{figure}

Since $\kappa$ scales generally as $\kappa=\kappa_0 t^\nu$,
where here $t=\frac{T}{T_C^{ext}}-1$ is the reduced temperature~\cite{Law84},
we estimated the correlation length exponent $\nu$
from the slope of a double logarithmic plot of the fitted $\kappa$ versus
$t$~(Fig.~\ref{Fig2}); the values obtained were $0.48\pm0.05\leqslant\nu\leqslant0.58\pm0.07$
when neglecting the crossover in $\nu$ and in the susceptibility exponent $\gamma$
at larger reduced temperatures~\cite{Tre98}. The range in $\nu$ is essentially attributable
to experimental uncertainties, e.g., the limited resolution
for smaller $t$. Note that we measure essentially the same value of $\nu$ 
and of the other critical exponents regardless of depth. Thus, the
value of the critical exponents are
\emph{depth-independent}.

Our values of $\beta$, $\nu$ and $\gamma_1$ ($0.96\pm0.13$ as reported
in Ref.~\onlinecite{Tre98}) all support tricritical behavior in the skin
layer for small $t$ when the tricritical point is approached along the $T$-axis.
Although the correlation length decreases with depth, our values
of $\nu$ and $\gamma_1$ (the subscript ``1'' refers to small $t$) compare
quite well with the theoretical values of $\nu=0.5$ and $\gamma_1=1$
obtained from the analysis of a metamagnet which yields mean field
exponents~\cite{Nel75,Law84}. They also compare well with other systems
presumed to be tricritical, e.g., $\nu=0.52\pm0.08$ and $\gamma_1=1.05\pm0.20$
for Nd${}_4$Cl~\cite{Yel74}.
\begin{figure}
 \centering
\includegraphics[width=0.45\textwidth]{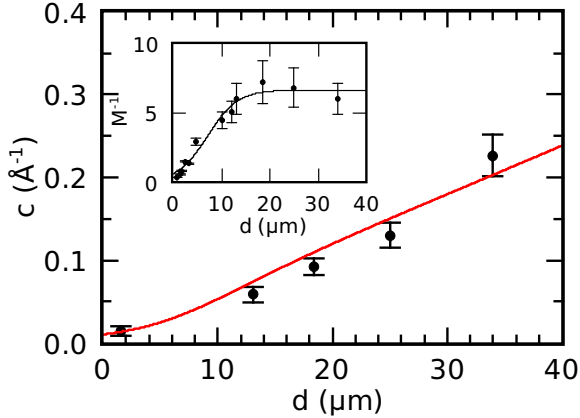}
\caption{\label{Fig3}(Color online) Proportionality factor $c$ (eq.~\ref{Scaling})
vs. depth $d$. The solid red line is the value of $c$ calculated according to
eq.~\ref{experc}. Inset: inverse of the mosaic spread (dimensionless) versus depth
$d$.}
\end{figure}

\section{Analysis and theoretical model}

According to the criteria
defined by Krivoglaz~\cite{Kri96}, the composition of the sample was
close enough to the tricritical point to observe tricritical behavior.
Although the experimental value of $\beta$ is smaller than the theoretically
expected value of 0.25, it is comparable with an earlier measurement and with
other tricritical systems if the influence of the two-phase region is
neglected~\cite{Tre00,BSch87}. Given this good agreement between the values
for the critical exponent, we believe that tricritical behavior is dominant
for small reduced temperatures.

Despite the fact that the critical exponent $\nu$ is depth-independent,
there is a depth dependence to the measured behavior of the correlation length,
or, equivalently, to the inverse correlation length $\kappa$.
To account for this depth-dependence,
we propose that the scaling law for
$\kappa$ be modified such that
\begin{equation}\label{Scaling}
 \kappa(d) =c\left( d\right) t^\nu\:,
\end{equation}
where the factor $c$ is dependent on the depth $d$.
To determine the function $c\left( d\right)$ we calculated the $y$-intercepts
of the fitted $\kappa$ vs.~$t$ plots shown in Fig.~\ref{Fig2}.
The resulting measured values for $c\left( d\right)$ are shown in Fig.~\ref{Fig3}.

These results indicate that, for the same change in $t^\nu$, the
inverse correlation length decreases faster for smaller depths. On the other hand,
the critical ordering being measured is presumably
occurring near the defect lines due to their strain fields~\cite{Dub79,Kor99},
which cause the appearance of ordered regions. We can then safely assume that the
density of ordered regions at any given depth is proportional to the density of
dislocation lines at the same depth. The mosaic spread gives us a measure of this
density, so we expect $c\left( d\right)$ to be proportional to the inverse of the
mosaic spread. However, it should be noted that since $\kappa(d)$ is calculated 
from the half width of the CDS profiles at depth $d$, it is an averaged measure.
In fact, the CDS profiles measure ordering throughout the skin-layer to
the depth that is probed. Thus, arguably, $c\left(d\right)$ should be proportional to
an integral average of the inverse of the mosaic spread $M$ from the surface to depth $d$:
\begin{equation}
 c\left(d\right)\propto\frac{1}{Z}\int_0^d\frac{1}{M\left(z\right)}\mathrm{d}z\:.
\end{equation}
The factor $Z$ has to account for the absorption of the X-rays along the path
through the material, so that it is, in itself, an integrated quantity:
\begin{equation}
 Z\left(d\right)=\int_0^{l\left(d\right)}\mathrm{e}^{-\mu x}\mathrm{d}x\:,
\end{equation}
where $\mu$ is the X-ray absorption coefficient and $l$ is the effective path into
the material, which, for a depth $d$ and a scattering angle $\vartheta$ is given by
\begin{equation}
 l\left(d\right)=\frac{2d}{\sin\vartheta}\:.
\end{equation}
We can then express $Z$ directly as
\begin{equation}
 Z\left(d\right)=\frac{1}{\mu}\left(1-\mathrm{e}^{-2\mu d/\sin\vartheta}\right)
\end{equation}
and, absorbing $\mu$ into the proportionality, give our final expression
for the experimentally measured $c$ as
\begin{equation}\label{experc}
  c\left(d\right)\propto\frac{\int_0^d\frac{1}{M\left(z\right)}\mathrm{d}z}{1-\mathrm{e}^{-2\mu d/\sin\vartheta}}\:.
\end{equation}
This quantity, calculated with average values of $\mu$ and $\sin\vartheta$
and using a sigmoid function fit for $M$,
is shown as a line in the main part of Fig.~\ref{Fig3}. The proportionality
constant used, dependent on the particular sample and the details of the
experiment, had a fitted value of $0.1\:\mathrm{m}^{-2}$. In the inset of the
same figure we show the depth dependence of the inverse of the mosaic spread.
Notice also that in this estimate for the experimentally measured $c$, 
$\mu$ and $\sin\vartheta$ only play a significant role for small depths, due to
their presence in a negative exponential.

\begin{figure}
 \centering
\includegraphics[width=0.45\textwidth]{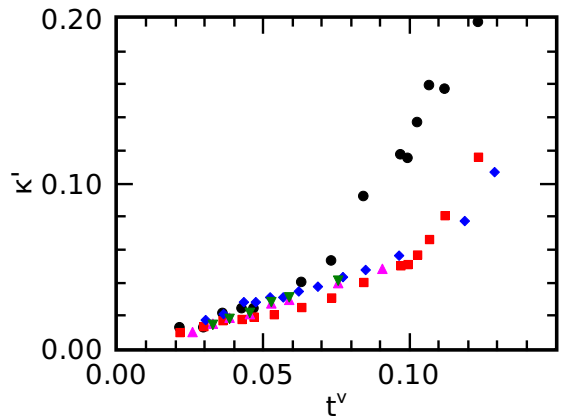}
\caption{\label{Fig4}(Color online) Inverse correlation scaling function
$\kappa'$ versus scaled
reduced temperature $t^\nu$, showing the data collapse. The black circles
correspond to 1.6~$\mu$m, the red squares correspond
to 13.1~$\mu$m, the blue diamonds to
18.4~$\mu$m, the pink upward triangles to 25~$\mu$m and the green downward triangles
to 34~$\mu$m.}
\end{figure}

With this treatment for $c\left( d\right)$,
it is possible to collapse the scattering data for
$\kappa$ onto a single scaling function
\begin{equation}
\kappa'=\frac{\kappa}{c\left( d\right)}
\end{equation}
as shown in Fig.~\ref{Fig4}.
Note that when $t^\nu$ goes to 0 the collapsed curves extrapolate to 0.
This corresponds to a vanishing inverse correlation length, or,
equivalently, to a diverging correlation length and thus
suggests that $T_C^{ext}\left( d\right) $ is the actual critical temperature,
or at least that it is not too far from the real critical temperature $\hat{T}_C\left( d\right)$.
The tails of the curves, for $t^\nu\gtrsim0.07$ correspond to
the critical region above the two-length scale crossover~\cite{Tre98}.

The collapse is strong evidence that the density of defects directly drives
the critical behavior in the material. The mechanism involved is very likely
to be the interaction with the stress and strain fields induced by the defects,
which is most probably also responsible for the crossover to different
values of the critical exponents and for the change in the order of the transition
between bulk and skin-layer.

\section{Conclusions}

In summary, we have observed a change in the order of the phase transition
in a V${}_2$H crystal, which is first-order in the pure bulk and continuous
in a skin layer that is several microns thick.
Defects, in the form of dislocation lines, exist in the skin layer and
are responsible for the change in the ordering behavior.
The density of those defects is inhomogeneously distributed and
decays continuously with depth from the surface.
Throughout the skin layer, near the critical temperature, the measured 
values of the critical exponents are those of a tricritical point.
Although the values of the critical exponents describing the scaling
behavior of thermodynamic functions in the skin layer do not depend
on depth, the coefficients of the power law function of $\kappa$ and,
presumably, of other thermodynamic functions, do show a depth
dependence. Focusing on the behavior of $\kappa$, 
we find that its depth-dependent scaling coefficient depends on an integral
average of the inverse defect density. Using this fact and postulating
a modified scaling law for the inverse correlation length, we are
able to collapse the measured data onto a single scaling function.
Thus, we are able to treat scattering measurements at different depths
in a single unified framework.

\begin{acknowledgments}
The authors would like to thank R.~Hempelmann for loading the crystal used
in these experiments and D.~Lott, H.~D.~Carstanjen, P.~C.~Chow, D.~De~Fontaine,
J.~W.~Cahn and R.~Barabash for help in the experiment or fruitful discussions. Furthermore,
we thank G.~Srajer and the beamline personnel at the SRI-CAT at the APS at the
Argonne National Laboratory for assistance during the experiment.
The work of CIDG and KEB was supported by the NSF through grant No.~DMR-0427538.
SCM gratefully acknowledges the support of the Texas Center for Superconductivity
of the University of Houston ($\mathrm{T_c}$SuH).
The Advanced Photon Source is supported by the U.S. DOE, BES-DMS, under contract
W-31-109-ENG-38.
\end{acknowledgments}

\end{document}